\documentclass[runningheads]{llncs}
\usepackage[T1]{fontenc}

\usepackage{graphicx}
\usepackage{float}
\usepackage[inline]{enumitem}
\usepackage[]{todonotes} %
\usepackage{csquotes}
\usepackage[style=lncs,giveninits=true,maxnames=3,mincitenames=1,maxcitenames=1,backend=biber,sortcites=true,natbib]{biblatex}
\usepackage{microtype}
\usepackage{booktabs}

\usepackage[hidelinks,bookmarks=false,final,breaklinks]{hyperref}
\usepackage{listings}
\usepackage{tabularx}
\usepackage{comment}

\usepackage[absolute]{textpos}
\usepackage{pgfplots}
\pgfplotsset{compat=1.18}
\usepackage{xcolor}
\usepackage{kitcolors}
\usepackage{pgf-pie}
\usepackage{subcaption}
\begin{document}
\title{Software Architecture Meets LLMs:\\A Systematic Literature Review}
\titlerunning{Software Architecture Meets LLMs: A Systematic Literature Review}

\author{Larissa Schmid\inst{1}\thanks{Work was conducted while affiliated with Karlsruhe Institute of Technology (KIT).}\orcidID{0000-0002-3600-6899} \and
Tobias Hey\inst{2}\orcidID{0000-0003-0381-1020} \and\\
Martin Armbruster\inst{2}\orcidID{0000-0002-2554-4501} \and
Sophie Corallo\inst{2}\orcidID{0000-0002-1531-2977} \and\\
Dominik Fuch\ss{}\inst{2}\orcidID{0000-0001-6410-6769} \and
Jan Keim\inst{2}\orcidID{0000-0002-8899-7081} \and\\
Haoyu Liu\inst{2}\orcidID{0009-0002-7676-5010} \and
Anne Koziolek\inst{2}\orcidID{0000-0002-1593-3394}
}
\institute{
KTH Royal Institute of Technology, Stockholm, Sweden\\
\email{lgschmid@kth.se}
\and
Karlsruhe Institute of Technology (KIT), Karlsruhe, Germany\\
\email{\{hey,martin.armbruster,sophie.corallo,dominik.fuchss,\\jan.keim,haoyu.liu,koziolek\}@kit.edu}
}
\authorrunning{Schmid et al.}

\maketitle              
\begin{abstract}
Large Language Models (LLMs) are used for many different software engineering tasks. In software architecture, they have been applied to tasks such as classification of design decisions, detection of design patterns, and generation of software architecture design from requirements. However, there is little overview on how well they work, what challenges exist, and what open problems remain.
In this paper, we present a systematic literature review on the use of LLMs in software architecture. We analyze 18 research articles to answer five research questions, such as which software architecture tasks LLMs are used for, how much automation they provide, which models and techniques are used, and how these approaches are evaluated. 
Our findings show that while LLMs are increasingly applied to a variety of software architecture tasks and often outperform baselines, some areas, such as generating source code from architectural design, cloud-native computing and architecture, and checking conformance remain underexplored. 
Although current approaches mostly use simple prompting techniques, we identify a growing research interest in refining LLM-based approaches by integrating advanced techniques. 

\end{abstract}

\section{Introduction}

Large language models (LLMs) are revolutionizing software engineering by offering exceptional capabilities in natural language understanding and generation.
These models can significantly enhance productivity by, e.g., automating code generation~\cite{adnan2025leveragingllmsdynamiciot}, helping in the development of cyber-physical systems \cite{p_shaukat_llm4cps} and digital twins~\cite{p_zhang_llm4DT,macias_icsa_DT}, analyzing logs~\cite{ase23_log_parsing,ase24_log_parsing}, and answering developing questions~\cite{llm4QAarch}.
Their ability to analyze vast amounts of data and identify patterns also helps in optimizing system performance and predicting potential issues early: 
LLMs can not only streamline workflows but also foster innovation and improve overall software quality.

Software architecture tasks also often require vast knowledge.
Consequently, there is an emerging synergy between software architecture and LLMs.
For example, LLMs have been explored for architecture tasks such as identifying design decisions~\cite{keim_a,soliman_exploring}, generating architecture designs from requirements~\cite{jahic_state}, and answering questions about architectural knowledge~\cite{soliman_do}.
Moreover, software architecture can be applied to developing LLM-based systems by providing reference architectures for different use cases~\cite{donakanti_reimagining,weber_fhgenie,lu_towards,shamsujjoha_swiss}.

Systematic literature reviews on the use of LLMs provide researchers and practitioners with comprehensive insights into current trends, challenges, and best practices.
These reviews help identify gaps in existing knowledge, guide future research directions, and inform evidence-based decision-making in the development and application of LLMs. 
However, reviews on the use of LLMs in the scope of software engineering mainly focus on testing~\cite{slr-llm4setesting} or code generation~\cite{zan2023llmsmeet}.
Most articles in existing literature reviews on LLM usage in general software engineering \cite{fan2023review,slr-llm4se} are not related to software architecture:
\citet{slr-llm4se} do not include works from the software architecture community, as the search keys did not include relevant terms (only "software design" could be considered related).
Even for design-related tasks, \citet{fan2023review} report that they did not find much work on LLM-based software design.
Different from code generation and tests, architecture tasks often affect higher level concerns and encounter data scarcity problems.
These disparities highlight the value of a comprehensive review of software architecture and LLMs.
To the best of our knowledge, no such review exists, making a systematic literature review particularly useful.

Therefore, in this paper, we conduct a systematic literature review of research papers at the intersection between LLMs and software architecture.
We formulate our research questions to derive insights into the current state-of-the-art in this field and about what is working well, challenges, and open questions.
From the software architecture side, we analyze the software architecture tasks targeted by these works and how the performance of LLMs is evaluated.
From the LLMs' side, we explore which LLMs are used and how they are optimized.
To provide insights into the path ahead for the synergy between LLMs and software architecture, we also analyze the discussed future work.
Moreover, we give an initial overview on envisioned reference architectures for developing LLM systems.
Following the methodology for systematic literature reviews on software engineering \cite{kitchenham2007guidelines,kitchenhamn2023guidelines}, we initially found 119 with our search strategy, of which we identify and analyze 18 relevant papers about LLMs and architecture.
We provide the complete data of this survey as supplementary material~\cite{supplementary-material}.

This literature review can benefit
\begin{enumerate*}[label=(\roman*)]
    \item software architecture researchers who want to apply LLMs in their architecture tasks and
    \item LLM-based systems developers who want to build their LLM systems with better architecture.
\end{enumerate*}

In the following, we first present our methodology to the review in \autoref{sec:approach}.
After that, \autoref{sec:findings} presents our findings on our RQs, and \autoref{sec:ttv} discusses threats to validity.
We further discuss our findings and outline future research directions in \autoref{sec:discussion}.
Finally, \autoref{sec:conclusion} concludes the paper.

\section{Methodology} \label{sec:approach}

This section describes the approach we followed to select, analyze, and evaluate relevant research on the intersection of software architecture and LLMs. We follow the methodology defined by Kitchenham et al.~\cite{kitchenham2007guidelines,kitchenhamn2023guidelines}. 
Therefore, our review process consists of three main phases: 1. Planning the review by formulating research questions of interest (\autoref{sec:approach:rqs}) and defining a search strategy (\autoref{sec:approach:search strategy}), 2. filtering the articles obtained by our search (\autoref{sec:approach:filtering}), and 3. analyzing the remaining relevant articles (\autoref{sec:approach:analysis}).

\subsection{Research Questions} \label{sec:approach:rqs}

Our review aims to provide an overview of the current applications of LLMs to software architecture research and vice versa, i.e., how software architecture research is applied to LLMs. We want to provide insight into what works well, what does not, and what challenges remain.

First, we investigate which software architecture tasks LLMs are used for (\textbf{RQ1}) to understand which tasks are already being researched and potentially solved, and which remain an open challenge.
To gain more detailed insight, we examine the degree of automation these approaches provide (\textbf{RQ1.1}), distinguishing between manual guidance, semi-automated, and fully automated methods.
Additionally, we assess whether LLMs are applied end-to-end or only to specific sub-tasks within the broader software architecture process (\textbf{RQ1.2}).

Since the LLMs' capabilities can vary significantly, our goal is to identify which LLMs are used in the reviewed studies (\textbf{RQ2}). 
This research question provides insight into the most commonly applied models and whether there is a preference for general-purpose or domain-specific LLMs in software architecture.

To understand how researchers tune LLM performance, we examine the techniques used to improve effectiveness (\textbf{RQ3}). 
Specifically, we investigate the used tuning techniques (\textbf{RQ3.1}) and prompt engineering strategies (\textbf{RQ3.2}).
RQ3 as well as RQ2 are based on the investigation by \citet{slr-llm4se}.

Evaluating the effectiveness of LLM-based approaches is crucial for understanding their practical applicability. Therefore, we explore how these approaches are evaluated (\textbf{RQ4}) by analyzing the evaluation methods used (\textbf{RQ4.1})~\cite{konersmann2022} and the specific metrics applied (\textbf{RQ4.2}).
Furthermore, we examine whether these methods outperform existing baselines (\textbf{RQ4.3}) and assess whether supplementary materials 
are provided (\textbf{RQ4.4}) to support reproducibility.

Finally, to gain insights into the future directions of LLM research in software architecture, we analyze what future work the authors of the reviewed studies suggest (\textbf{RQ5}). Identifying open challenges and proposed research directions helps outline the next steps to advance LLM applications in this domain.

\subsection{Search Strategy} \label{sec:approach:search strategy}

We extracted a search query based on our goal to provide an overview of the applications of LLMs in software architecture and vice versa.
Therefore, one part of our query is the keyword \textit{software architecture} that has to be within the article for us to regard it as relevant. 
Moreover, the article has to contain a keyword related to LLMs.
As not all articles may use the same keyword related to their usage of LLMs, we included several different terms the article has to contain at least one of, including the currently most popular models: \textit{"LLM" OR "language model" or "language models" OR "generative AI" OR "bert" OR "GPT" OR "Llama" OR "Transformer"}.

We use this search string to search for articles in 25 top software engineering conferences and journals, such as ICSE, ASE, ICSA, ECSA, TSE, TOSEM, by using Google Scholar and defining them as sources of the publication.
We provide the complete list of venues as part of our supplementary material~\cite{supplementary-material}. 
For the most closely related conferences, namely ICSA and ECSA, we modify the search string not to need to contain the term \textit{software architecture}, as the scope of the conference already implies that the article is related to software architecture.
Moreover, we also include companion proceedings from these two conferences.
This search leads to 119 articles.

\subsection{Filtering of Results} \label{sec:approach:filtering}

Based on our initial search, resulting in 119 articles, we filter the results to make sure the articles are actually relevant to our survey.
First, we check if they contain the term \textit{software architecture} as part of the article and find that 44 articles only mention it as part of the references, e.g., when citing an article from a software architecture conference.
Next, we assess if an article is a full article from the research track of the respective conference -- except for ICSA and ECSA, where we also consider contributions from the companion proceedings -- and if it conducts research on the topic of software architecture and LLMs. We distribute this step among the team of authors and refer to the ICSA scope of topics in the call for papers for determining if the article is in the scope of software architecture as \textit{inclusion criteria}.
Notably, this implies the \textit{exclusion criteria} that domain and UML models like class and activity diagrams are excluded. 
Moreover, we \textit{exclude} research related to only design patterns as opposed to architectural patterns. 
We filter 12 articles based on not being full articles, and 15 more because they are not related to software architecture and LLMs.
We filter one additional article because it is a survey article; therefore, it only discusses existing research and does not propose a new approach.
We notice that two more articles, while presenting slightly different ideas, show the same evaluation. 
Therefore, we subsume them into one article.

\newpage
\subsection{Analysis of articles} \label{sec:approach:analysis}

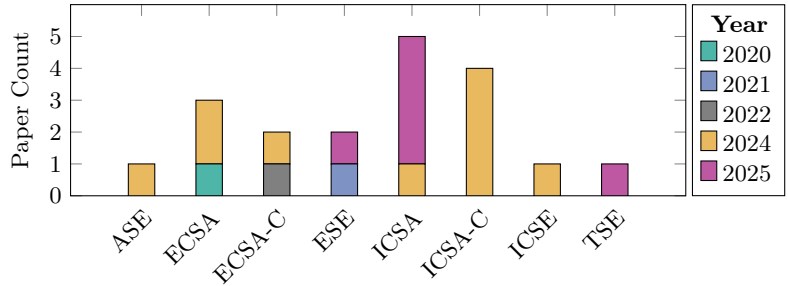
\begin{figure}
    \centering
    \begin{tikzpicture}
        \begin{axis}[
            ybar stacked,
            symbolic x coords={ASE, ECSA, ECSA-C, ESE, ICSA, ICSA-C, ICSE, TSE},
            xtick=data,
            ymin=0, ymax=6,
            ytick={0,1,2,3,4,5},
            ylabel={Paper Count},
            legend pos=north east,
            legend style={at={(1.01,0.5)}, anchor=west},
            enlarge x limits=0.15, 
            width=.8\textwidth,
            height=4.125cm, 
            xticklabel style={rotate=45, anchor=north east}
        ]
        \addlegendimage{empty legend}

        \addplot[fill={kit-green70}] coordinates {(ASE,0)(ECSA,1)(ECSA-C,0)(ESE,0)(ICSA,0)(ICSA-C,0)(ICSE,0)(TSE,0)}; 
        \addplot[fill={kit-blue70}] coordinates {(ASE,0)(ECSA,0)(ECSA-C,0)(ESE,1)(ICSA,0)(ICSA-C,0)(ICSE,0)(TSE,0)}; 
        \addplot[fill={kit-gray50}] coordinates {(ASE,0)(ECSA,0)(ECSA-C,1)(ESE,0)(ICSA,0)(ICSA-C,0)(ICSE,0)(TSE,0)}; 
        \addplot[fill={kit-orange70}] coordinates {(ASE,1)(ECSA,2)(ECSA-C,1)(ESE,0)(ICSA,1)(ICSA-C,4)(ICSE,1)(TSE,0)}; 
        \addplot[fill={kit-purple70}] coordinates {(ASE,0)(ECSA,0)(ECSA-C,0)(ESE,1)(ICSA,4)(ICSA-C,0)(ICSE,0)(TSE,1)}; 

        \addlegendentry{\hspace{-.2cm}\textbf{Year}}
        \addlegendentry{2020}
        \addlegendentry{2021}
        \addlegendentry{2022}
        \addlegendentry{2024}
        \addlegendentry{2025}

        \end{axis}
    \end{tikzpicture}
    \caption{Distribution of relevant articles regarding venue and year of publication.}
    \label{fig:article-distribution}
\end{figure}

After filtering the results, we end up with \textbf{18 unique and relevant articles}. 
\autoref{fig:article-distribution} shows their distribution by venue and year of publication. 
While the first article was already published in 2020 at ECSA, there was only one publication in the following years, and there was no publication in 2023. 
However, 2024 shows a steep increase with 10 articles, five of them being part of companion proceedings. 
In 2025, five articles have already been published, further indicating that the upward trend in publications will continue.
Most articles (14/18) are published at ICSA, ECSA, or in their companion proceedings. 
This is not surprising as both conferences are the most closely related to the topic of our review.

We then extracted relevant data from the articles to answer our research questions outlined above (cf. \autoref{sec:approach:rqs}).

\section{Findings} \label{sec:findings}

In the following, we present our findings from the analysis of articles on our research questions.

\subsection{RQ1: Software Architecture Tasks} \label{sec:findings:rq1}

Our first research question investigates the software architecture tasks, the degree of automation of the approaches, and how LLMs are utilized in these tasks.

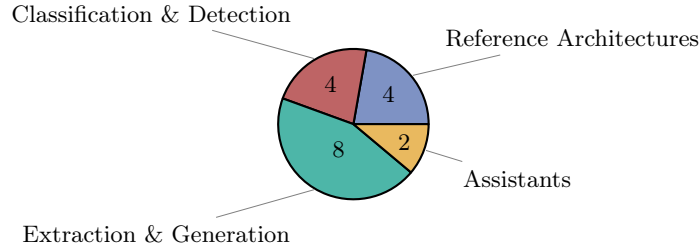
\begin{figure}
    \centering
    \begin{tikzpicture}
        \pie[radius=1, sum=18, text=pin, color={kit-blue70, kit-red70, kit-green70, kit-orange70}]{
            4/Reference Architectures,
            4/Classification \& Detection,
            8/Extraction \& Generation,
            2/Assistants
        }
    \end{tikzpicture}
    \caption{Distribution of Tasks Utilizing LLMs (n=18)}
    \label{fig:rq1-categories}
\end{figure}

We identified four main categories of software tasks related to software architecture that utilize LLMs:
Reference Architectures, Classification \& Detection, Extraction \& Generation, and Assistants.
We provide an overview of the distribution of these categories in \autoref{fig:rq1-categories}.

Reference architectures cover domains such as self-adaptive systems \cite{donakanti_reimagining}, chatbots with LLMs \cite{weber_fhgenie}, and agents \cite{lu_towards,shamsujjoha_swiss}.
We discuss them shortly in \autoref{sec:discussion}. 

Classification and detection tasks include classifying tactics in code \cite{keim_does}, design decisions \cite{keim_a}, and identifying design decisions in mailing lists \cite{soliman_exploring}.
LLMs are also used as classifiers in traceability link recovery tasks \cite{keim_recovering}.

Extraction and generation tasks involve extracting design rationales \cite{zhao_drminer}, architecture component names \cite{fuchss_enabling}, design structures from code \cite{fang_a}, and mining design discussions \cite{mahadi_conclusion}.
Regarding generation, creating architecture decision records \cite{dhar_can}, software architecture designs from requirements \cite{jahic_state}, and architecture components for FaaS \cite{arun_llms} are application scenarios for LLMs.
Also, the generation of module descriptions and text embeddings for model-to-code mappings \cite{johansson_mapping} are part of this category.

Assistant systems focus on question-answering about architectural knowledge \cite{soliman_do} and aiding in selecting, assessing, and capturing better design decisions \cite{diaz-pace_helping}.

Most of the works (71~\%) use LLMs in an automated fashion.
The two approaches that build assistants or chatbots are semi-automated, as they require user interaction.
In the remaining categories, only two further studies are classified as semi-automated, while the rest are fully automated.
The semi-automation is related to either providing adaptable infrastructure components for identifying types of architectural design decisions rather than fully automation \cite{soliman_exploring} or requiring the user to define and enter prompts themselves \cite{jahic_state}.

Whether the LLM is used to solve a subtask or the entire task is mixed across the studies.
While 64~\% of studies use LLMs end-to-end, 36~\% of studies use them for subtasks.
We observed the following subtasks for the non-assistant categories:
Classification tasks \cite{soliman_exploring}, generation of descriptions or embeddings \cite{johansson_mapping}, extraction of component names \cite{fuchss_enabling}, and generation of explanations \cite{fang_a}.
Moreover, one of the assistants \cite{diaz-pace_helping} uses LLMs for multiple subtasks like suggesting patterns, ranking, assessment of decisions, and generation of architecture decision records.

\subsection{RQ2: Which Large Language Models are used?} \label{sec:findings:rq2}
\begin{figure}
    \centering
    
    
    
    \centering
    \begin{tikzpicture}
        \begin{axis}[
                ybar stacked,
                xtick=data,
                xticklabel style={rotate=0, anchor=north},
                axis x line*=bottom,
                axis y line=none,
                bar width=1em,
                font=\footnotesize,
                nodes near coords,
                every node near coord/.append style={font=\tiny, align=center, text=black},
                enlarge x limits=0.24,   
                xtick distance = 6em,
                enlarge y limits=0.1, 
                symbolic x coords={
                        BERT, T, GPT, Llama, Other
                    },
                axis line style={draw=none},
                tick style={draw=none},
                ymin=2,
                width=1.1\linewidth,
                height=25em
            ]
            \node[anchor=west,font=\tiny,xshift=0.33em] at (axis cs:GPT,0.5) {GPT-2};
            \node[anchor=west,font=\tiny,xshift=0.33em] at (axis cs:GPT,2) {GPT-3};
            \node[anchor=west,font=\tiny,xshift=0.33em] at (axis cs:GPT,6) {GPT-3.5};
            \node[anchor=west,font=\tiny,xshift=0.33em] at (axis cs:GPT,11.5) {GPT-3.5 turbo};
            \node[anchor=west,font=\tiny,xshift=0.33em] at (axis cs:GPT,17.5) {GPT-4};
            \node[anchor=west,font=\tiny,xshift=0.33em] at (axis cs:GPT,22) {GPT-4 turbo};
            \node[anchor=west,font=\tiny,xshift=0.33em] at (axis cs:GPT,23.5) {GPT-4o mini};
            \node[anchor=west,font=\tiny,xshift=0.33em] at (axis cs:GPT,24.5) {GPT-4o};
            
            \node[anchor=west,font=\tiny,xshift=0.33em] at (axis cs:BERT,1.5) {BERT};
            \node[anchor=west,font=\tiny,xshift=0.33em] at (axis cs:BERT,4) {SentenceBERT};
            \node[anchor=west,font=\tiny,xshift=0.33em] at (axis cs:BERT,5.5) {DistilBERT};
            \node[anchor=west,font=\tiny,xshift=0.33em] at (axis cs:BERT,6.5) {CodeBERT};
            \node[anchor=west,font=\tiny,xshift=0.33em] at (axis cs:BERT,7.5) {UniXCoder};
            
            \node[anchor=west,font=\tiny,xshift=0.33em] at (axis cs:T,0.5) {T5};
            \node[anchor=west,font=\tiny,xshift=0.33em] at (axis cs:T,1.5) {Flan-T5};
            \node[anchor=west,font=\tiny,xshift=0.33em] at (axis cs:T,2.5) {T0};

            \node[anchor=west,font=\tiny,xshift=0.33em] at (axis cs:Llama,0.5) {Llama 2};
            \node[anchor=west,font=\tiny,xshift=0.33em] at (axis cs:Llama,1.5) {Llama 3.1};
            \node[anchor=west,font=\tiny,xshift=0.33em] at (axis cs:Llama,3) {CodeLlama};
            
            \node[anchor=west,font=\tiny,xshift=0.33em] at (axis cs:Other,0.5) {ULMFiT};
            \node[anchor=west,font=\tiny,xshift=0.33em] at (axis cs:Other,1.5) {DeepSeek-V2.5};
            \node[anchor=west,font=\tiny,xshift=0.33em] at (axis cs:Other,2.5) {Artigenz-Coder};
            \node[anchor=west,font=\tiny,xshift=0.33em] at (axis cs:Other,3.5) {CodeQwen1.5};
            
            \addplot[draw=black,fill=kit-green70] coordinates {(GPT,1) (BERT,3) (T,1) (Llama,1) (Other,1)}; 
            \addplot[draw=black,fill=kit-blue70] coordinates {(GPT,2) (BERT,2) (T,1) (Llama,1) (Other,1)}; 
            \addplot[draw=black,fill=kit-orange70] coordinates {(GPT,6) (BERT,1) (T,1) (Llama,2) (Other,1)}; 
            \addplot[draw=black,fill=kit-red70] coordinates {(GPT,5) (BERT,1) (T,0) (Llama,0) (Other,1)}; 
            \addplot[draw=black,fill=kit-purple70] coordinates {(GPT,7) (BERT,1)}; 
            \addplot[draw=black,fill=kit-cyan70] coordinates {(GPT,2)}; 
            \addplot[draw=black,fill=kit-lightgreen70] coordinates {(GPT,1)}; 
            \addplot[draw=black,fill=kit-brown70] coordinates {(GPT,1)}; 
            
        \end{axis}
    \end{tikzpicture}
    
    \caption{Distribution of Used Models Grouped by Their Base Approach}
    \label{fig:stacked_bar_chart_models_grouped}
\end{figure}
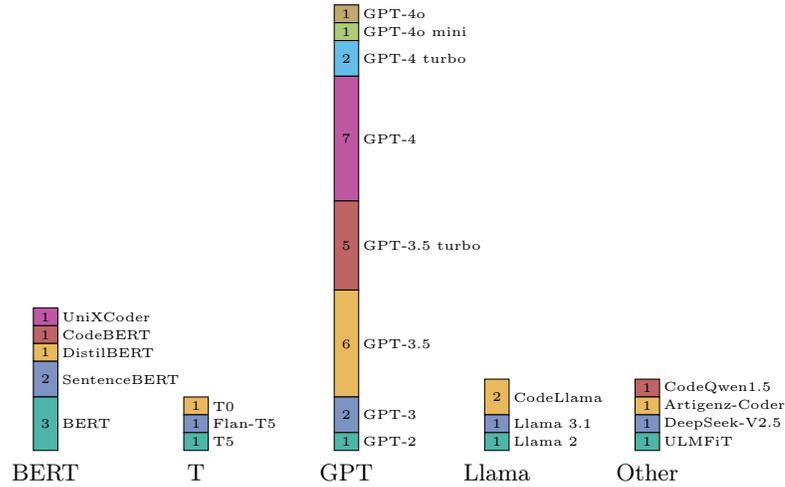
To give a more detailed insight into the capabilities of the LLMs used in the studies, we analyzed the distribution of the respective models.
This distribution is displayed in \autoref{fig:stacked_bar_chart_models_grouped}.
In total, 23 different models were used, which we grouped according to the base approach they derived from.
However, all but ULMFiT are based on the Transformer architecture~\cite{vaswani_attention_2017}.
We included ULMFiT~\cite{howardUniversalLanguageModel2018} anyways, as it is the first language model that introduces transfer learning with task-specific fine-tuning and can thus be seen as the direct predecessor of the current LLMs.
The first observation one can make is that most of the used models (73~\%) are only using the decoder-part of the Transformer architecture (GPT-, Llama-, and DeepSeek-based models).
Encoder-only models (BERT-based and ULMFiT) are used in 21~\% of the studies and encoder-decoder models (T-based) only in 7~\% of the cases.
This was expected, as since the release of GPT-3 the auto-regressive (decoder-only) LLMs surpassed the other variants in most SE tasks \cite{slr-llm4se}.
This also aligns with the distribution of the models over time: until 2024, solely encoder-only models were used (two times BERT~\cite{keim_does,keim_a} and one time ULMFiT~\cite{mahadi_conclusion}).
However, as GPT-3 was released in Mai 2020, the adoption of the decoder-only LLMs in the software architecture community was slower than in other SE areas~\cite{slr-llm4se,slr-llm4setesting}.
The most recent versions of GPT-based models (GPT-4o and later) and Llama-based models (Llama 3.1 and later), as well as the DeepSeek-based models (DeepSeek-V2.5 and Artigenz-Coder), were only used in the 2025 publications~\cite{fuchss_enabling,arun_llms}.
This trend might be underestimated, as our study only includes data until March 2025.

\subsection{RQ3: Optimization Techniques} \label{sec:findings:rq3}

We observed a clear distinction in tuning approaches based on the type of model.
Encoder models were consistently fine-tuned across studies, emphasizing the need for task-specific adaptation due to their transformer-based masked language modeling pre-training.
Fine-tuning allows researchers to tailor the model to software architecture tasks by training it on domain-specific data.

In contrast, decoder models like those from the GPT family were predominantly utilized through prompting techniques rather than fine-tuning, likely due to accessibility and cost constraints.

\begin{figure}[t]
    \centering
    \begin{subfigure}[b]{0.48\textwidth}
        \centering
        \begin{tikzpicture}
            \pie[radius=1, sum=20, text=pin, color={kit-blue70, kit-red70, kit-green70, kit-yellow70, kit-orange70}]{
                14/zero-shot , 3/few-shot, 1/cot, 1/rag, 1/template
            }
        \end{tikzpicture}
        \caption{Prompting Techniques (n=20)}
        \label{fig:rq3-rq4-prompting-techniques}
    \end{subfigure}
    \hfill
    \begin{subfigure}[b]{0.5\textwidth}
        \centering
        \begin{tikzpicture}
            \pie[/tikz/every pin/.style={align=center},radius=1, sum=22, rotate=150, text=pin, color={kit-blue70, kit-red70, kit-green70, kit-yellow70, kit-orange70, kit-purple70}]{
                9/technical\\experiment, 6/benchmark, 4/case study, 1/controlled\\\vspace*{.5em}experiment, 1/data science, 1/interviews
            }
        \end{tikzpicture}
        \caption{Evaluation Methods (n=22)}
        \label{fig:rq3-rq4-evaluation-methods}
    \end{subfigure}
    \caption{Overview of used prompting techniques and used evaluation methods}
    \label{fig:rq3-rq4}
\end{figure}
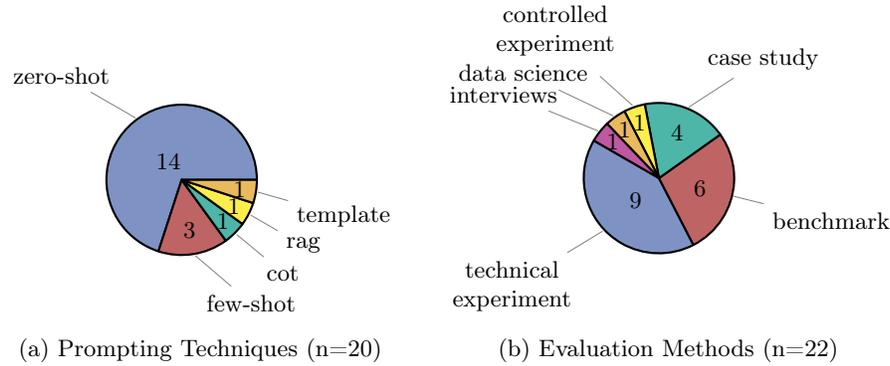

\autoref{fig:rq3-rq4-prompting-techniques} shows the overview of prompting techniques used.
We found that researchers most commonly employed zero-shot prompting (70~\% of used techniques).
This aligns with the general usability of LLMs, as zero-shot prompting allows direct application without further training.
This can also mean that the pre-trained LLMs encode enough knowledge for many software architecture tasks.

Few-shot prompting was used less frequently (15~\%), suggesting that providing examples is not necessarily required for software architecture tasks.

More advanced prompt engineering strategies were rarely applied, which could indicate an area for future exploration.
Chain-of-Thought prompting was used only in one study despite its potential to improve reasoning-based tasks.
Retrieval-Augmented Generation was also applied only once, indicating that integrating external knowledge sources is not yet a common practice in this domain.
Similarly, template-based prompting appeared in a single instance, suggesting that structured prompt design is underexplored for software architecture tasks.

\subsection{RQ4: Evaluation of Approaches} \label{sec:findings:rq4}

For the evaluation of LLM-based approaches in software architecture tasks (RQ4.1), the most common methods were technical experiments and benchmarking, followed by case studies (cf.~\autoref{fig:rq3-rq4}).
Technical experiments were the dominant evaluation method, used in 64~\% of studies (cf.~\autoref{fig:rq3-rq4-evaluation-methods}).
Benchmarking was conducted in 43~\% of studies, often involving comparisons with traditional or state-of-the-art approaches.
Case studies were used in 29~\% of studies, offering qualitative insights into real-world applications.
Other evaluation methods each only appeared once, including data science-based validation, interviews, and controlled experiments.

Looking into RQ4.2, the evaluation of the LLM-generated outputs employed both traditional performance metrics and text-generation metrics.
Traditional performance metrics (e.g., precision, recall, F\textsubscript{1}-score) were frequently applied to measure the correctness of LLM-generated outputs.
Text generation metrics, which are used to assess the quality of generated content, include BLEU (Bilingual Evaluation Understudy) and BERTScore.
BLEU was adopted by three studies (21~\%; i.e., \cite{zhao_drminer, dhar_can, arun_llms}).
BERTScore, which evaluates semantic similarity using contextual embeddings, appeared in one study (i.e., \cite{dhar_can}).

A key question in assessing the effectiveness of LLMs for software architecture is whether they outperform existing approaches (RQ4.3).
Among the fourteen studies analyzed, nine included a comparison to other approaches.
Five studies did not compare their methods to a baseline, limiting their ability to demonstrate relative effectiveness.
In cases where a comparison was conducted, LLM-based solutions consistently outperformed the baseline in six studies.
Two studies showed mixed results: \citet{mahadi_conclusion} demonstrated better results within-dataset, but worse across, and \citet{keim_a} performed better than the baselines according to the F\textsubscript{1}-score, but showed lower precision in some cases. Another study~\cite{keim_does} was not able to outperform the baseline.
However, these results suggest a generally positive impact of LLMs on software architecture tasks.
While these results highlight the potential of LLMs, the lack of baseline comparisons in one-third of the studies indicates a need for more rigorous benchmarking to establish their practical advantages.

RQ4.4 tackles reproducibility as it is a crucial aspect of scientific research, enabling independent verification of results.
Nearly all studies provided some form of supplementary material, such as datasets, source code, or implementation details.
However, two works proposing reference architectures did not include additional materials, possibly due to the conceptual nature of their contributions.
This suggests a strong commitment to reproducibility within the field, though improvements in providing accessible and well-documented supplementary materials could further enhance transparency.

\subsection{RQ5: Future Work} \label{sec:findings:rq5}

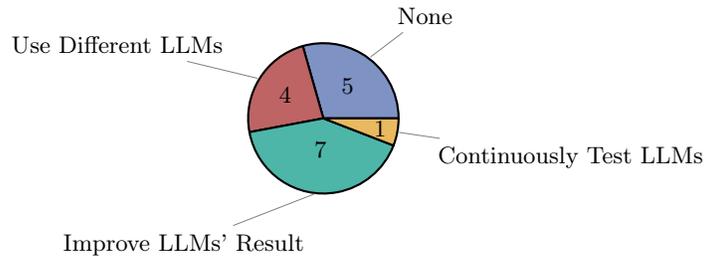
\begin{figure}
    \centering
    \begin{tikzpicture}
        \pie[radius=1, sum=17, text=pin, color={kit-blue70, kit-red70, kit-green70, kit-orange70}]{
            5/None,
            4/Use Different LLMs,
            7/Improve LLMs' Result,
            1/Continuously Test LLMs
        }
    \end{tikzpicture}
    \caption{Number of occurrences for different categories of future work (n=17).}
    \label{fig:findings:rq5-distribution-pie}
\end{figure}

Regarding RQ5, we considered the future work mentioned and related to LLMs. In total, five papers (36~\%) do not report on future work \cite{zhao_drminer,johansson_mapping,keim_a,soliman_exploring,keim_recovering}, while nine papers (64~\%) give a short outlook \cite{fuchss_enabling,dhar_can,soliman_do,keim_does,fang_a,jahic_state,diaz-pace_helping,mahadi_conclusion,arun_llms} (cf. \autoref{fig:findings:rq5-distribution-pie}).


In these nine papers, future work aims to expand the papers' results in three different directions. First, four studies want to use different LLMs for testing \cite{soliman_do}, with integrated reasoning \cite{fuchss_enabling}, with code support \cite{keim_does}, or with multi-modal capabilities \cite{diaz-pace_helping}. Second, seven studies want to improve the LLMs' results, in general, \cite{soliman_do} or with specific approaches. This includes a preprocessing or refinement of the input \cite{fuchss_enabling,diaz-pace_helping}, adding more context to the input \cite{fuchss_enabling,dhar_can}, applying different techniques (e.g., RAG) to the LLM \cite{dhar_can,fang_a,diaz-pace_helping,arun_llms}, or fine-tuning the LLM \cite{dhar_can,mahadi_conclusion}. Third, in one study, the authors plan to test LLMs for software architecture tasks continuously \cite{jahic_state}.

\section{Threats to Validity} \label{sec:ttv}

In the following section, we discuss threats to validity~\cite{Wohlin2012}. 
One threat involves not finding all relevant articles due to our search strategy and query employed. 
We mitigated this threat by evaluating different queries and the relevance of papers found with them beforehand. We also checked if the results included relevant papers we knew of as a gold standard~\cite{dieste2009developing}. 
Another threat is the misclassification of articles. We need to extract information from the articles to answer our research questions, making it necessary to understand them correctly. 
All authors have expertise in the research field of our review, and papers were assigned based on their knowledge of the respective areas. Moreover, we discussed any issues that arose among the complete team of authors, ensuring consistent and accurate classification.

\section{Discussion}
\label{sec:discussion}

In the following, we discuss our findings from \autoref{sec:findings} and identify future research directions. 

\paragraph{Software Architecture Tasks.} 
Our study shows the diverse applications of LLMs in software architecture, with tasks falling into four main categories (cf. \autoref{sec:findings:rq1}): reference architectures, classification \& detection, extraction \& generation, and assistants. 
Examining the 14 articles on LLMs for software architecture, we found that most of them propose automated approaches that use LLMs end-to-end, suggesting that LLMs are capable of addressing complete architectural tasks. 

Besides the 14 articles covering the applications of LLMs in software architecture, we also found four articles concerning the application of software architecture to LLMs. 
They propose reference architectures for incorporating LLMs into different domains, such as self-adaptive systems, chatbot frameworks, and autonomous agents. 
By structuring interactions between LLMs and external systems, software architecture enables more robust and adaptable applications, stressing how software architecture research can not only use LLMs but also benefit them. 

Surprisingly, we found only one work generating source code for architectural components using LLMs~\cite{arun_llms}. 
Also, there is only one paper regarding cloud-native computing and architecture~\cite{arun_llms}, indicating a potential avenue for further research in this regard. 
We found no articles regarding evaluating quality aspects of software architecture, such as evolvability, and architecture conformance checking.
Both could be addressed in future research, e.g., by building on works identifying architectural patterns~\cite{fang_a} and design rationales~\cite{zhao_drminer} from code.

\paragraph{Usage of LLMs.}

Most approaches (73~\%) rely on decoder-only models (\autoref{sec:findings:rq2}), particularly GPT-based variants, reflecting their dominance in recent research. 
This trend of using mostly decoder-only, GPT-based LLMs can also be observed in the broader software engineering context~\cite{slr-llm4se,slr-llm4setesting}.
However, there is also no consensus for a specific variant~\cite{slr-llm4se}. 
Fine-tuning was common for encoder models, whereas decoder models were primarily used via prompting (\autoref{sec:findings:rq3}), with zero-shot prompting being the most frequent strategy (70~\%). This also aligns with findings in the software testing context\cite{slr-llm4setesting}, where zero-shot prompting is also the most used strategy, followed by few-shot prompting.
In the broader software engineering context, \citet{slr-llm4se} found that few-shot prompting was the most commonly employed strategy, followed by zero-shot prompting. 
All surveys, including ours, show that advanced prompting techniques, like Chain-of-Thought and Retrieval-Augmented Generation, are only rarely used. 
Exploring whether these techniques can enhance approaches used for software architecture tasks is a question for future research. 

\paragraph{Evaluation of Approaches.}
Evaluation methods were mainly technical experiments and benchmarking, with F\textsubscript{1}-score being the most commonly used metric (\autoref{sec:findings:rq4}). 
While most studies showed LLMs outperforming baselines, around one-third lacked comparative evaluation to a baseline.
This indicates a need for more rigorous validation to demonstrate the added benefits of utilizing LLMs.
However, nearly all studies provide supplementary material, enabling further insight into the approaches and results.

\paragraph{Future Work.}
Future research directions mentioned by the authors of the studies include testing different LLMs, refining input strategies, and integrating advanced techniques such as retrieval-augmented generation (RAG) and fine-tuning. 
These findings suggest that while LLMs offer significant potential for software architecture tasks and outperform baselines, it is a multi-dimensional problem to apply them in a way that ensures the best results.

\paragraph{The Future of LLMs in Software Architecture.}
Our findings indicate that the current body of published research on this topic is relatively limited.
This is consistent with the review by \citet{fan2023review} that characterized LLM-based design as an open research direction. 
Yet, there seems to be emerging research, as shown by the number of workshop publications and at this year's ICSA. 
One of the reasons for the comparatively low number of papers in software architecture as opposed to other software engineering disciplines could be that the capabilities of LLMs were not sufficient to perform software architecture tasks until then:
The three studies from before 2024 that utilize encoder-only models were not able to demonstrate consistent improvements of their approaches over the baselines~\cite{keim_a,keim_does,mahadi_conclusion}. 
This also illustrates the need for the continuous evaluation of both LLMs and proposed approaches for software architecture tasks:
Given the fast-paced development of LLM technology, future research should consider strategies for ongoing assessment and adaptation of models in software architecture contexts.

\section{Conclusion} \label{sec:conclusion}

In this paper, we present a systematic literature review on the usage of LLMs in software architecture and vice versa. Analyzing 18 relevant articles, we find that LLMs are mainly applied to tasks within Classification \& Detection, Extraction \& Generation, Assistants, and Reference Architecture, also showing how software architecture can be applied to LLMs. 
Most studies used decoder-only models, specifically GPT-based variants. 
While the number of articles covering software architecture and LLMs is increasing, the current body of research on this topic is relatively limited, and some areas, such as generating source code from architectural design, cloud-native computing and architecture, and checking conformance, are not well covered. 

Most studies evaluate LLM-based approaches through technical experiments and benchmarking, often using performance metrics like the F\textsubscript{1}-score. However, about one-third of the studies do not compare their results to a baseline, showing that stronger comparative evaluation is needed. 
Additionally, most studies use basic prompting techniques, while more advanced methods, such as Chain-of-Thought prompting and Retrieval-Augmented Generation, are rarely explored.  

Our results suggest that future research should focus on improving LLM performance by refining input strategies, using more advanced prompting techniques, and regularly testing model capabilities. As LLMs continue to improve, their role in software architecture will likely grow, making continuous evaluation important to ensure their reliability and usefulness.
As LLMs continue to evolve and more research emerges, this review should be repeated soon to keep up with new developments and trends. 
\section{Data Availability} \label{sec:data-availability}

We provide the complete data of this survey as supplementary material~\cite{supplementary-material}, i.e., the considered venues, the complete list of found articles, the filtered articles, and the classification of the examined articles. 
\section*{Acknowledgments}
This work was funded by Core Informatics at KIT (KiKIT) of the Helmholtz Assoc. (HGF), by KASTEL Security Research Labs, and the German Research Foundation (DFG) - SFB 1608 - 501798263.

\printbibliography[heading=bibintoc]
\end{document}